# Parallel Firewalls on General-Purpose Graphics Processing Units

Manoj Singh Gaur and Vijay Laxmi
Kamal Chandra Reddy, Ankit Tharwani, Ch.Vamshi Krishna, Lakshminarayanan.V
*Department of Computer Engineering*
*Malaviya National Institute of Technology, Jaipur, India*
{gaurms,vlaxmi}@mnit.ac.in
{kamal359, ,vamshi990 ,strings2ln}@gmail.com

*Abstract* -**Firewalls use a rule database to decide which packets will be allowed from one network onto another thereby implementing a security policy. In high-speed networks as the inter-arrival rate of packets decreases, the latency incurred by a firewall increases. In such a scenario, a single firewall become a bottleneck and reduces the overall throughput of the network.A firewall with heavy load, which is supposed to be a first line of defense against attacks, becomes susceptible to Denial of Service (DoS) attacks[LMR].Many works are being done to optimize firewalls.This paper presents our implementation of different parallel firewall models on General-Purpose Graphics Processing Unit (GPGPU). We implemented the parallel firewall architecture proposed in [2] and introduced a new model that can effectively exploit the massively parallel computing capabilities of GPGPU.**

## I. INTRODUCTION

A firewall is a part of a computer system or network that is designed to block unauthorized access while permitting authorized communications [3]. It is a device or set of devices that is configured to permit or deny network packets based upon a set of rules. Firewalls can be implemented in either hardware or software, or a combination of both. Firewalls are used either on a host or on an edge device(gateway/router). Firewalls are frequently used to prevent unauthorized Internet users from accessing private networks connected to the Internet, especially intranets. All packets entering or leaving the intranet pass through the firewall, which inspects each packet and blocks those that do not meet the specified security criteria. The basic components in building a firewall: network policy, advanced authentication mechanisms, packet filtering and application gateway.

Packet filtering involves finding a matching rule for a packet, based on its metadata, from a set of rules. A packet filter is a software which looks at the header of packets as they pass through, and decides the fate of the entire packet. Packet filter may decide to DROP the packet as if it had never received it, or ACCEPT the packet i.e., to let the packet go through, or something take some action as specified to it[4].

Under Linux, packet filtering is built into the kernel as part its networking stack. Netfilter is a framework for packet mangling, outside the normal Berkeley socket interface. Each protocol defines hooks(IPv4 defines 5) which are well-defined points in a packet's traversal of that protocol stack. At each of these points, the protocol will call the netfilter framework with the packet and the hook number. Also, parts of the kernel can register to listen to the different hooks for each protocol. So when a packet is passed to the netfilter framework, it checks to see if any routine has registered for that protocol and hook; if so, the registered routines get to examine the packet in order, then discard the packet, allow it to pass, or ask netfilter to queue the packet for userspace[5].

The *iptables* tool inserts and deletes rules from the kernel's packet filtering table.Packet filtering rules consist of :Insertion Point (INPUT/OUTPUT/FORWARD), Insertion Order, one or more matching criteria, single target to specify action (ACCEPT / DROP /QUEUE). '*iptables*' uses *default deny* and *first match* policy. Hence more specific rules should be on the top of the ruleset. In Linux, packet filtering is done sequentially i.e., packets are processed in order one after the other.

## II. RELATED WORK

Amongst the components of a firewall mentioned, packet filtering is performance critical since it checks every ingress/egress packets. With new generation internet applications and increase in the number of hosts con- necting to the internet the bandwidth requirements of organisations increases frequently. As the traffic from/to the intranet goes high, the load on the firewall i.e.

number of packets to be filtered per unit time increases. Unless the firewall scales up to the network speed, the firewall becomes a bottleneck in the network.

Many optimization techniques are employed to increase throughput and/or decrease per packet latency of firewalls. Some of the current optimization strategies include algorithmic optimization [7], implementing filtering in FPGA/ASIC [8] and parallel firewalls [2]. Depending on the specific requirements one or more optimisataion techniques can be used to build a firewall for high-speed networks.

Parallel firewall architecture were introduced by Erin F. et al.[2]. Parallelization can greatly enhance the performance of network firewalls by offering scalability to reduce processing loads. Parallelizing firewalls can be implemented by either dividing the traffic or the workload across an array firewall nodes .All firewall nodes should be able to work in parallel. Thus, a firewall node can be a node in a cluster or a processor in a multi-core system.

## III. CUDA PROGRAMMING MODEL

A GPU is a highly parallel, multi-threaded, many core dedicated processor attached to the graphics card, capable of offloading 2D or 3D graphics rendering from the CPU. General purpose operations on GPUs are made possible by recent developments. Because of the parallel structure, high computational power and memory bandwidth, the general purpose GPUs are more effective than the general purpose CPUs for many applications.

CUDA[6] is a parallel programming model that leverages the computational capacity of the GPUs for non-graphics applications. It is a set of extensions to the C programming language and provides support that applications can use both the CPU and the GPU for serial and parallelareas of the code respectively, the GPU thus being a co-processor.

The instructions to be executed on the device are encapsulated within a kernel function and the GPU is capable of launching a large number of threads operating on identical instructions, but on different sets of data, in a Single Instruction Multiple Data (SIMD) manner. A thread block (or block) is a group of threads (upto 512 threads executing concurrently) that work together efficiently by means of fast sharing of data and synchronization to coordinate memory accesses. Each thread has a local memory(register spill space). Each block has shared memory, visible to all the threads within the block and within the span of the block. During execution, each block is assigned to a multiprocessor, each of which consists of on-chip memory including a set of local 32-bit registers, a shared memory, a read-only constant cache and a read-only texture cache.

## IV. IMPLEMENTATION ON CUDA

Under Linux, packet filtering is built into the kernel (as a kernel module, or built right in). From the kernel tree of linux-2.6.32.14 the file /net/ipv4/netfilter/ip_tables.c contains the packet matching code. This code is executed for every packet entering or leaving the network.

In our work we are using CUDA to implement packet filtering that will execute in parallel on GPGPU.Unfortunately Nvidia does not provide any drivers which allow the direct control of their cards from within the Linux kernel. This restriction forces all interactions with their cards to originate from userspace processes, making the use of CUDA services from within the kernel a challenge. To overcome this restriction, we are queuing packets to Userspace, process them in the GPGPU and reinject back to the kernel networking stack with action decided from filtering.

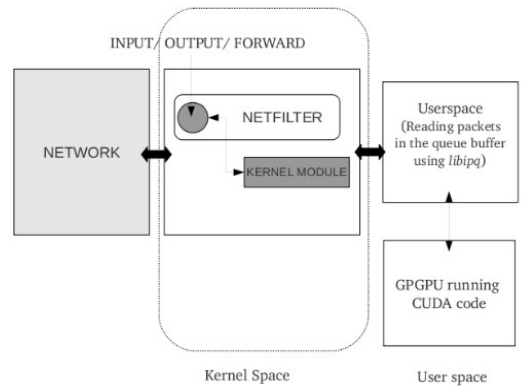

Fig. 1. Queuing packets to userspace

The firewall testbed created uses a set of rules to perform filtering. Each rule contains 5-tuple: source IP address, destination IP address, source port number, destination port number, protocol and specifies the action (accept/deny). If a particular filed is not specified for some rule will match packets for all values of that field. All the packets across the network are compared with the list of rules and the action is taken accordingly. Packet filtering rules consist of Insertion Point (INPUT/ OUTPUT/ FORWARD) , Insertion Order one or more matching criteria, Single target to specify action (ACCEPT / DROP /QUEUE) .

## A. Data-Parallel Firewall

Using the terminology of parallel computing, a design that distributes the data(packets) across the firewall nodes is considered data parallel. A data parallel firewall system consists of multiple identical firewalls connected in parallel. Each $i^{th}$ firewall node, needs a local policy (rule set) $R_i$ which is a duplicate of the complete security policy allowing it to act independently. Arriving packets are then distributed across the firewall nodes such that only one firewall node processes any given packet. Therefore different packets are processed in parallel and all packets are compared to the entire security policy.

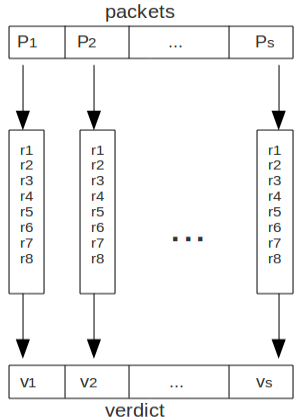

Fig. 2. Data-Parallel Firewall

The data parallel approach is based on the model specifed in [2] with some modifications that are implementation specific to CUDA programming model. A set of packets are accumulated, processed in parallel by different threads on GPGPU and reinjected back to network stack based on the verdict issued.

## B. Function-Parallel Firewall

The function parallel approach is based on the model specifed in [2] with some modifications that are implementation specific to CUDA programming model. A function parallel design consists of an array of firewall nodes. In this design, arriving traffic is duplicated to all firewall nodes. Each firewall node 'i' employs a part of the security policy. After a packet is processed by each firewall node 'i', the result of each $R_i$ is sent to the calling program runing on the CPU.. To ensure no more than one firewall node determines action on a packet only the sequential code can execute an action on a packet.

A packet is processed by multiple threads simultaneously with each thread checking a part of the ruleset. Packets are copied from host to device global memory. Part of rule database is kept in device shared memory. Verdict array is copied from device to host after all the threads completes execution.

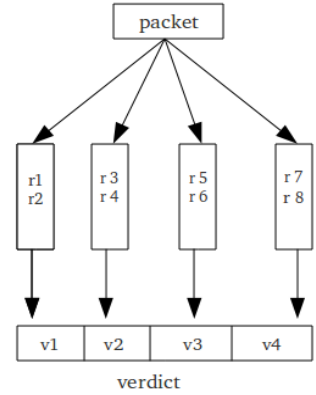

Fig. 3. Function-Parallel Firewall

## C. Hybrid Model

The data-parallel and function-parallel designs does not fully exploit the computational capabilities of CUDA architecture. Data-parallel design increases the throughput of the firewall. Function-parallel design reduces per packet processing delay. Using massively parallel computing capabilities of GPGPU we have acheived a better model with increased throughput and reduced packet processing delay.

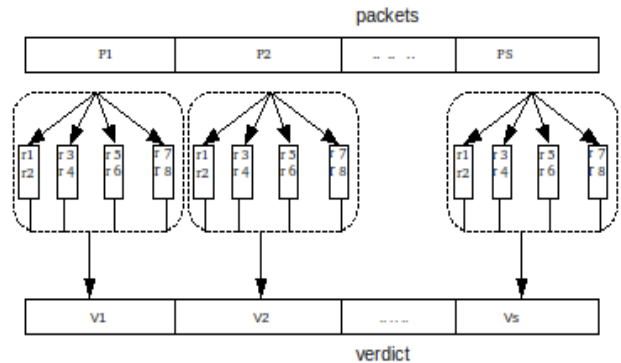

Fig. 4. Hybrid Model

In Hybrid Model, a set of packets are accumulated and each packet is processed in parallel by a different block on GPGPU. Within a block, a packet is processed by multiple threads simultaneously with each thread checking a part of the ruleset. Rule database is kept in device global memory. Verdict array is copied from device to host after all the threads completes execution.

## IV. RESULTS AND ANALYSIS

We ran our models on a node that has 2.80 GHz Dual Core processor with 512 MB RAM. We used a nVidia Quadro FX1700 GPGPU with 512MB global memory and 32 cores. The applications were first executed on CPU. These were compared with the timings from the parallel models executed on CPU+GPGPU.

### A. Limitations for Data-Parallel Design

Accesses to constant memory are cached. But this caching does not improve performance since the cached data is not used later. Overhead of copying the whole rulest to shared memory eclipses the improvement achieved by reduction in contenetion for global memory.

Average Packet Delay (global memory)   : **15.625 µs**
Average Packet Delay (constant memory) : **15.531 µs**
Average Packet Delay (shared memory)   : **21.156 µs**

### B. Analysis of Function-Parallel Design

Because of the CUDA overhead in copying verdict array to/from the device memory and kernel launch for every packet makes function-parallel model inefficient. However, the effect of different kernel launch configurations on the packet delay has been studied.

For a given ruleset size, as the number of threads increase rules per thread decreases. The effect of different kernel launch configurations can be explained by considering,
- Overhead in copying rules to shared memory ( rules per thread )
- Contention to access data from shared memory (threads per block)

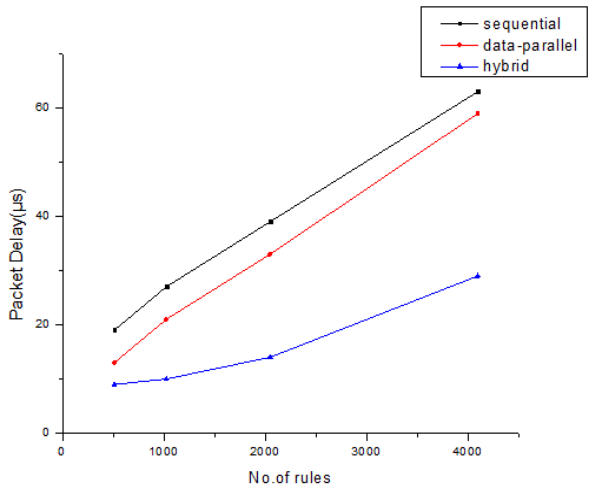

Fig. 5. Packet delay as the number of rules increases. Parallel designs consisted of 64 threads

Simulations are performed to test the latency[9] of various modeles of Parallel Firewalls. Each is compared with the conventional sequential approach. Traffic is generated using HPING3. Traffic used for performance analysis is homogeneous. Packet processing time is calculated in each case by varying the number of firewall nodes as well as the ruleset size.

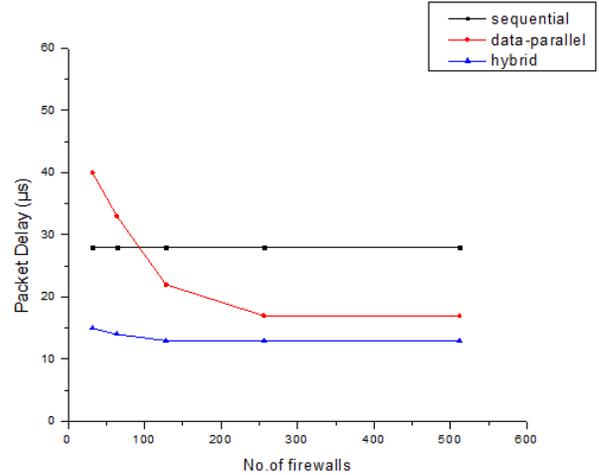

Fig. 6. Packet delay as the number of firewalls increases. Ruleset consisted of 2048 rules

## V. CONCLUSIONS AND FUTURE WORK

We have explored the use of GPGPUs for Packet Filtering in Firewalls. The three prototypes we developed for GPGPU kernels for implementing Parallel Firewalls suggest that it is possible to use GPGPUs in the design of High-Performance Network Systems. The positive results indicate that GPGPUs can be part of next-generation high-speed networks. There is also room for further optimizations, which we haven't explored fully. There remain a few pending issues with our performance analysis which is to be done with a traffic simulating a real-time behaviour. We also intend to port all the code to OpenCL so as to experiment on different architectures and compare architecture specific code like CUDA with its OpenCL counterparts. Current parallel models are stateless. A stateful parallel firewall will pose new challenges to implement. CUDA kernel launches can be made asynchronous to reduce overall packet processing delay.

## VI. REFERENCES

[1] K.Salah et al., "Resiliency of Open-Source Firewalls against Remote Discovery of Last-Matching Rules", In Proceedings of the Security of Information and Networks 2009.

[2] E.W.Fulp. Parallel firewall designs for high-speed networks. In Proceedings of the 25 th IEEE International Conference on Computer Communications, pp1-4. INFOCOM 2006.

[3] D.Newman,"Benchmarking terminology for firewall performance" , RFC 2647, August 1999.

[4] R. Russell and H. Welte, "Linux Packet Filtering HOWTO".

[5] R. Russell and H. Welte, "Linux Netfilter Hacking HOWTO".

[6] www.nvidia.com/object/cuda/home /

[7] S. Suri and G. Varghese, "Packet Filtering in high speed networks," in Proceedings of the Symposium in Discrete Algorithms, 1999, pp.969-970.

[8] G.S. Jedhe, A.Ramamoorthy, and K.Varghese, "A Scalable High Throughput Firewall in FPGA", In Proceedings of the 16th International Symposium on Field-Programmable Custom Computing Machines, 2008.

[9]R.Hickman,D.Newman and T.Martin, "Benchmarking Methodology for Firewall Performance" RFC 3511, April 2003.